\documentclass[aps,letterpaper,twocolumn,prl,showpacs,nofootinbib]{revtex4-1}
\usepackage{amssymb}
\usepackage{amsmath,bm}
\DeclareMathOperator{\sech}{sech}
\usepackage[pdftex]{graphicx,color}
\usepackage[english]{babel}
\usepackage{hyperref}
\usepackage{enumitem}
\usepackage{epstopdf}
\usepackage{tikz}

\newcommand*\circled[1]{\tikz[baseline=(char.base)]{

            \node[shape=circle,draw,inner sep=1pt, fill = yellow] (char) {\textcolor{blue}{\small{#1}}};}}
\begin{document}

\title{Coexistence and interactions between nonlinear states with different polarizations in a monochromatically driven passive Kerr resonator}

\author{Alexander U. Nielsen$^{1,2}$}
\email{anie911@aucklanduni.ac.nz}
\author{Bruno Garbin$^{1,2}$}
\author{St\'ephane Coen$^{1,2}$}
\author{Stuart G. Murdoch$^{1,2}$}
\author{Miro Erkintalo$^{1,2,}$}
\email{m.erkintalo@auckland.ac.nz}
\affiliation{$^1$Department of Physics, University of Auckland, Auckland 1010, New Zealand}
\affiliation{$^2$The Dodd-Walls Centre for Photonic and Quantum Technologies, New Zealand}

\begin{abstract}
We report on experimental observations of coexistence and interactions between nonlinear states with different polarizations in a passive Kerr resonator driven at a single carrier frequency. Using a fiber ring resonator with adjustable birefringence, we partially overlap nonlinear resonances of two orthogonal polarization modes, achieving coexistence between different nonlinear states by locking the driving laser frequency at various points within the overlap region. In particular, we observe coexistence between temporal cavity solitons  and modulation instability patterns, as well as coexistence between two nonidentical cavity solitons with different polarizations. We also observe interactions between the distinctly polarized cavity solitons, as well as spontaneous excitation and annihilation of solitons by a near-orthogonally polarized unstable modulation instability pattern. By demonstrating that a single frequency driving field can support coexistence between differentially polarized solitons and complex modulation instability patterns, our work sheds light on the rich dissipative dynamics of multimode Kerr resonators. Our findings could also be of relevance to the generation of multiplexed microresonator frequency combs.

\end{abstract}

\maketitle

Dissipative solitons are localized structures that emerge in far-from-equilibrium nonlinear systems due to mechanisms of self-organization~\cite{akhmediev_dissipative_2005,coullet_localized_2002}. They manifest themselves in a myriad of different contexts, such as biology~\cite{tlidi_localized_2014,davydov_solitons_1977}, chemistry~\cite{kaufman_localized_1972,kyoung_pattern_1993}, hydrodynamics~\cite{russell_report_1844, fauve_solitary_1990}, and nonlinear optics~\cite{tlidi_localized_1994,barland_cavity_2002,ackemann_fundamenals_2009,sich_observation_2012,grelu_dissipative_2012,garbin_topological_2015}. In contrast to the more familiar solitons of conservative systems~\cite{akhmediev_solitons_1997}, dissipative solitons typically correspond to unique attractors of an underlying dynamical system, and their characteristics are consequently singularly determined by the parameters of the system~\cite{prigogine_selforg_1977}.

Temporal cavity solitons (CSs) are a particular breed of dissipative  solitons that manifest themselves in coherently-driven passive nonlinear optical resonators~\cite{wabnitz_suppression_1993,leo_temporal_2010, erkintalo_Kerrmedia_2016}. They are pulses of light, able to recirculate indefinitely in a resonator without changes in shape or energy. They have attracted significant attention in the context of high-Q microresonators~\cite{herr_temporal_2014} due to their role in the generation of broadband coherent frequency combs~\cite{rivas_dynamics_2014,webb_experimental_2016,weiner_frequency_2017,pfeiffer_octave-spanning_2017,palomo_microresonator_2017,yang_sync_2018,yi_imaging_2018}, and their fundamental characteristics and dynamics have been extensively investigated using macroscopic fiber ring resonators~\cite{leo_dynamics_2013, jang_ultraweak_2013, jang_observation_2014, jang_temporal_2015, luo_spontaneous_2015, anderson_observations_2016}.

As dissipative solitons, CSs are inherently unique attractors of the underlying dynamical system. This implies that, while several CSs can simultaneously coexist~\cite{leo_temporal_2010, jang_all-optical_2017}, all of them are expected to exhibit identical characteristics. Recent studies have shown~\cite{lucas_spatial_2018, weng_heteronuclear_2019}, however, that this expectation can fail when two (or more) fields with different frequencies drive the resonator: each driving field can then sustain CSs with distinct characteristics. In particular, researchers have used polychromatic driving fields to achieve coexistence between CSs associated with different spatial or polarization mode families of a microresonator~\cite{lucas_spatial_2018}, achieving compact sources of dual- and even triple-combs.

Interestingly, \emph{even a monochromatic driving field} can allow for the coexistence of distinct CS states, provided the driving field simultaneously excites several modes of the resonator. Such a situation has been theoretically argued~\cite{hansson_scs_2015} and experimentally demonstrated~\cite{anderson_coexistence_2017} to arise when the Kerr nonlinear phase shifts are sufficiently large so that neighbouring resonances of the same mode family partially overlap. One may similarly envision that, if the nonlinear resonances of two \emph{different} (spatial or polarization) mode families overlap, a single monochromatic driving field can simultaneously support nonidentical CSs associated with the two different modes. Averlant et al. have indeed theoretically predicted that a birefringent resonator driven with a suitably polarized monochromatic field could engender coexistence between two differentially polarized CSs with different peak powers and temporal durations~\cite{averlant_coexistence_2017}. Such coexistence has not, however, been experimentally observed.

In this Letter, we report on the first experimental observations of the coexistence of distinct nonlinear states associated with different polarization modes in a \emph{monochromatically} driven passive Kerr resonator. In addition to coexistence between two distinct CSs, we also observe coexistence between CSs and modulation instability (MI) patterns (both stable and unstable), and importantly, resolve complex interactions between the different nonlinear structures. Besides corroborating prior theoretical predictions~\cite{averlant_coexistence_2017,suzuki_theoretical_2018}, our experiments show that monochromatically-driven multimode Kerr resonators can display rich and previously unexplored dissipative dynamics that could be of applied relevance to the generation of multiplexed microresonator frequency combs~\cite{lucas_spatial_2018}. More generally, our results may help better understand the dynamics of dissipative localized states manifesting themselves in different two-component vectorial systems~\cite{williams_fast_1997,lecaplain_polarization_2013,marconi_vectorial_2015,averlant_vector_2016}.

We consider a ring resonator made of a single-spatial-mode waveguide with anomalous dispersion and weak birefringence. The resonator admits two mode families that correspond to the two orthogonal principal polarization modes of the waveguide. We assume the resonator is driven with a monochromatic laser polarized such that both modes can be simultaneously excited. In the limit of high finesse, the evolution of the slowly-varying intracavity electric field envelopes $E_{1,2}$ of each polarization mode is described by two coupled mean-field equations similar to the celebrated Lugiato-Lefever equation (LLE)~\cite{averlant_coexistence_2017,lugiato_spatial_1987}. In dimensionless form, the equations read~\cite{wang_universal_2017}:
\begin{align}
\begin{split}
\frac{\partial E_{1,2}(t,\tau)}{\partial t} &= \left[\vphantom{\frac{\partial^2}{\partial\tau^2}} -1 + i\left(|E_{1,2}|^2 + B|E_{2,1}|^2 - \Delta_{1,2}\right) \right. \\
  &\qquad \left. {}+i \frac{\partial^2}{\partial\tau^2} \right] E_{1,2} + S_{1,2}.
  \label{eq:1}
\end{split}
\end{align}

Here, $t$ is a slow time variable that describes the evolution of $E_{1,2}$ on a scale of the cavity photon lifetime, while $\tau$ is a corresponding fast time that describes the envelopes' temporal profile over a single round trip. The terms on the right-hand side of Eqs.~\eqref{eq:1} describe, respectively, the resonator losses, the self-phase modulation, the cross-phase modulation (XPM)~\cite{agrawal_nonlinear_1989}, the resonator frequency detuning, the group-velocity dispersion, and the coherent driving. The normalized driving field amplitudes $S_{1,2}$ are given by $S_{1} = S\cos(\chi)$ and $S_{2} = S\sin(\chi)$, where the driving ellipticity $\chi$ determines the projection of the total normalized driving field amplitude $S = \sqrt{X}$ ($X$ is the normalized driving power) into each of the polarization modes.

Equation~\eqref{eq:1} neglects linear mode coupling, group-velocity mismatch, coherent four-wave mixing mode interactions, and higher-order dispersive and nonlinear terms. With these assumptions, the two fields are coupled exclusively via XPM, with the coefficient $B$ describing the strength of that coupling. (In the simulations that will follow, we set $B = 1.3$~\cite{wang_universal_2017}.) The good agreement between our experiments and simulations justifies the use of this simplified model.

\begin{figure}[ht]	
		\includegraphics[width=\linewidth]{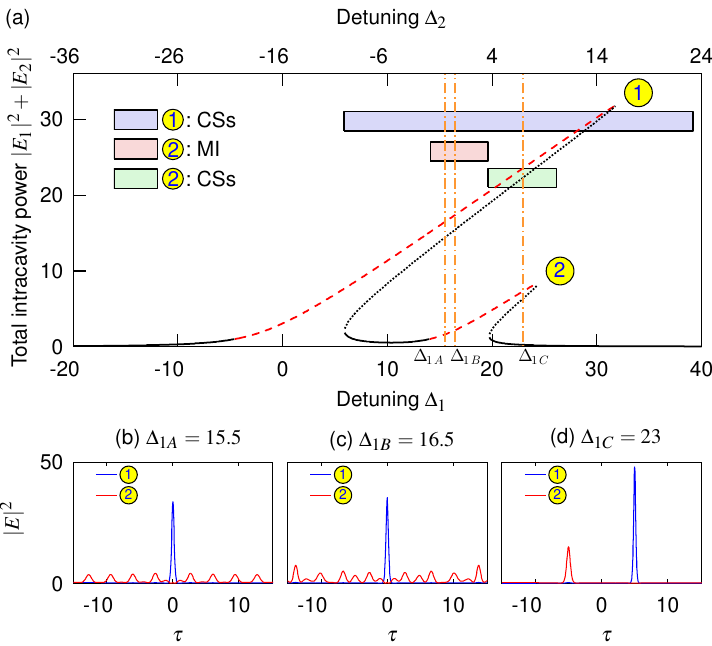}
		\caption{Cavity resonances and examples of coexisting nonlinear states calculated for $X = 40$, $\delta\Delta = 16$, $\chi = 0.15\pi$ and $B = 1.3$. (a) Total intracavity power ($|E_1|^2+|E_2|^2$) corresponding to the cw steady-state solutions of Eqs.~\eqref{eq:1}. Solid black lines correspond to stable solutions, black dotted lines to unstable solutions, and the red dashed lines represent MI unstable solutions. (b)--(d) Field profiles for each polarization mode at selected detunings [orange dashed-dotted vertical lines in (a)], obtained from numerical simulations of Eqs.~(\ref{eq:1}). Coexistence between: (b) a stable MI pattern and a CS, $\Delta_{1A} = 15.5$; (c) an unstable MI pattern and a CS, $\Delta_{1B} = 16.5$; (d) two different CSs, $\Delta_{1C} = 23$.}
		\label{fig:1}
\end{figure}

Due to the birefringence of the resonator, the two polarization modes are associated with different resonance frequencies. Accordingly, the frequency detunings $\Delta_{1,2}$ between the monochromatic driving laser and each of these resonances is in general different, i.e., $\Delta_{1} \neq \Delta_{2}$. We denote the frequency separation of the resonances as $\delta\Delta = \Delta_{1} - \Delta_{2}$ and assume $\Delta_{1} > \Delta_{2}$ [see Fig.~\ref{fig:1}(a)].

The coexistence of different nonlinear states depends non-trivially on the cavity parameters~\cite{averlant_coexistence_2017}. Through extensive simulations, we find that uneven driving ($\chi\neq \frac{\pi}{4}$) and large resonance separation ($\delta\Delta \gg 1$) are favourable for experimental observation. Accordingly, in the experiments that will follow, we set $\delta\Delta = 16$, $X = 40$ and $\chi = 0.15\pi$, such that polarization modes \circled{1} and \circled{2} are strongly and weakly driven, respectively. [Note that $\delta\Delta \gg 1$ justifies the omission of linear mode coupling and coherent four-wave mixing terms in Eqs.~\eqref{eq:1}.] Figure~\ref{fig:1}(a) shows the predicted total intracavity power $|E_1|^2+|E_2|^2$ for these parameters as $\Delta_{1}$ (and hence $\Delta_2 = \Delta_1-\delta\Delta$) is varied, obtained by solving the continuous wave (cw) steady-state solutions of Eqs.~\eqref{eq:1}. As can be seen, the differentially tilted resonances overlap, suggesting they can be simultaneously excited with a monochromatic driving field.

The regions where different nonlinear states coexist can be qualitatively estimated based on the range of CS existence in the absence of XPM coupling. [Simulations show that this scalar estimation is accurate (albeit not exact) for our parameters, but becomes increasingly unreliable as the resonance separation $\delta\Delta\rightarrow 0$.] The blue shaded rectangle in Fig.~\ref{fig:1}(a) represents the range of detunings over which CSs can manifest themselves for polarization mode \circled{1} in the scalar approximation. This range overlaps with the regions where polarization mode \circled{2} is expected to sustain MI patterns (shaded red rectangle) or CSs (shaded green rectangle). We find that coexisting nonlinear states manifest themselves in the overlap regions. Indeed, in Figs.~\ref{fig:1}(b)--(d), we show field profiles for each polarization mode at selected detunings [indicated with orange vertical lines in Fig.~\ref{fig:1}(a)], obtained via numerical simulation of Eqs.~\eqref{eq:1}. At $\Delta_{1A} = 15.5$, a CS in polarization mode \circled{1} coexists with a near-periodic and stable MI pattern in polarization mode \circled{2} [Fig.~\ref{fig:1}(b)]. When increasing the detuning to $\Delta_{1B} = 16.5$, the MI pattern in mode \circled{2} becomes more aperiodic and unstable (fluctuating as a function of slow time $t$), yet continues to coexist with the CS in mode \circled{1} [Fig.~\ref{fig:1}(c)]. At $\Delta_{1C} = 23$, two nonidentical CSs belonging to different polarization modes coexist [Fig.~\ref{fig:1}(d)]. Note that, while the nonlinear states are predominantly polarized along one of the polarization modes, they exhibit a small component along the orthogonal polarization mode due to XPM.

\begin{figure}[b]	
		\includegraphics[width=\linewidth]{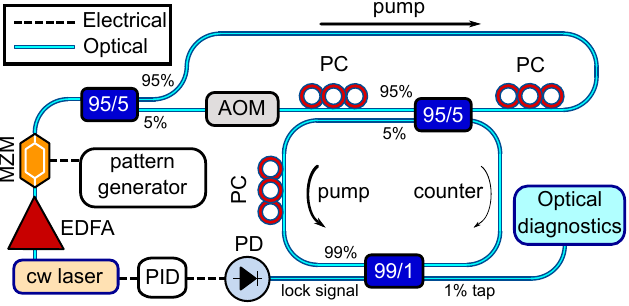}
		\caption{Schematic illustration of our experimental setup. MZM, Mach-Zehnder amplitude modulator; EDFA, Erbium-doped fiber amplifier; PC, polarization controller; PD, photodetector; AOM, acousto-optic modulator; PID, proportional-integral-derivative controller.}
		\label{fig:2}
\end{figure}

For experimental demonstration, we used an 85-m-long fiber-ring resonator (see Fig.~\ref{fig:2}) made out of standard single-mode fiber with group velocity dispersion and nonlinear coefficients of $\beta_{2} = -20 ~\mathrm{ps^2km^{-1}}$ and $\gamma = 1.2~\mathrm{W^{-1}km^{-1}}$, respectively. The fiber is laid in a ring configuration and closed on itself with a 95/5 coupler. The resonator also includes a 99/1 tap-coupler for monitoring the intracavity dynamics, as well as a polarization controller (PC) that allows us to systematically adjust the total birefringence of the cavity, and hence control $\delta\Delta$. The resonator has a measured finesse of $\mathcal{F} \approx 40$, and we synchronously drive it with flat-top nanosecond pulses carved from a 1550 nm narrow linewidth cw laser. A PC before the input coupler is used to adjust the amount of power projected into each polarization mode, i.e., to control $\chi$.

To observe coexistence between different nonlinear states, we lock the cavity detuning at different values using the method demonstrated in~\cite{nielsen_dm_2018}. In short, a PID-controlled feedback loop keeps the power level of a low intensity counter-propagating frequency shifted signal fixed, which in turn locks the detuning. At the 1 \% tap-coupler cavity output, we use a PC and a polarizing beam-splitter to separate the intensities of the two orthogonal polarization modes of the cavity for individual observation. Two optical spectrum analyzers are used to measure the spectral features of the different nonlinear states, while real-time dynamics are captured using $10~\mathrm{GHz}$ amplified photodiodes combined with a 40 $\mathrm{GSa/s}$ oscilloscope. Finally, CSs are excited by mechanically perturbing the resonator.

\begin{figure}[b]	
		\includegraphics[width=\linewidth]{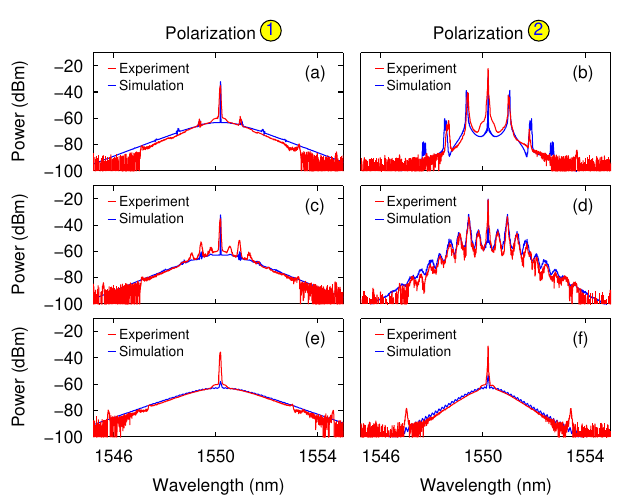}
		\caption{Experimentally measured (red curves) and numerically simulated (blue curves) spectra of the intracavity field in polarization modes 1 and 2, at detunings similar to those used in the simulations in Fig.~\ref{fig:1} (i.e. $\Delta_{1A}$, $\Delta_{1B}$ and $\Delta_{1C}$). The spectra are indicative of coexistence between: (a)--(b) a near-periodic and stable MI pattern and a CS, (c)--(d) an aperiodic and unstable MI pattern and a CS, (e)--(f) two nonidentical CSs.}
		\label{fig:3}
\end{figure}

Red curves in Fig.~\ref{fig:3} show experimentally measured, polarization-resolved spectra at cavity detunings similar to those used in Figs.~\ref{fig:1}(b)--(d). Also shown are numerically simulated spectra (blue curves) corresponding to the temporal profiles in Figs.~\ref{fig:1}(b)--(d). Figures~\ref{fig:3}(a) and~\ref{fig:3}(b) show spectra measured at a detuning where simulations predict [cf. Fig.~\ref{fig:1}(b)] CSs to coexist with a stable MI pattern. The spectrum of polarization mode \circled{2} indeed displays strong spectral sidebands (primary combs) characteristic of a near-periodic and stable MI pattern [Fig.~\ref{fig:3}(b)], while the spectrum of polarization mode \circled{1} shows the $\sech^2$ shape typical for a CS [Fig.~\ref{fig:3}(a)]~\cite{coen_universal_2013}. Note how XPM coupling results in the appearance of small sidebands on the CS's spectrum at the positions of the primary MI components. As the detuning is slightly increased, the strong spectral sidebands of the MI pattern broaden due to the aperiodicity of the pattern [Fig.~\ref{fig:3}(d)], in good agreement with corresponding simulations. Yet, the coexistence with the CS spectrum is still maintained on the orthogonal polarization [Fig.~\ref{fig:3}(c)].

By increasing the detuning lock point further, we are able to sustain coexistence between two distinct CSs. Indeed, Figs.~\ref{fig:3}(e) and~\ref{fig:3}(f) show that, at $\Delta_1 \approx 23$, spectra measured along both polarization modes exhibit the characteristic $\sech^2$ shape. The spectral widths measured along the two polarizations axes are noticeably different, as expected based on the different detunings experienced by the corresponding CSs~\cite{coen_universal_2013}. Of course, this also suggests the solitons have different temporal durations and peak powers [see Fig.~\ref{fig:1} (d)].

Additional temporally resolved measurements (not shown) reveal that the CSs are trapped to amplitude inhomogeneities on the opposite edges of our driving pulses~\cite{hendry_spontaneous_2018}. This observation can be readily understood in terms of the group-velocity mismatch between the two polarization modes. Specifically, because the two CSs travel at slightly different group velocities, synchronizing the driving pulses to the round trip time of one soliton would give rise to a large synchronization mismatch relative to the other, and hence cause the latter soliton to fall off the driving pulse. To minimize synchronization mismatch, we carefully adjust the pump pulse repetition time to be in between the repetition times of the two solitons, allowing for their trapping at the opposite edges of the pump pulse~\cite{hendry_spontaneous_2018}.

\begin{figure}[b]	
		\includegraphics[width=\linewidth]{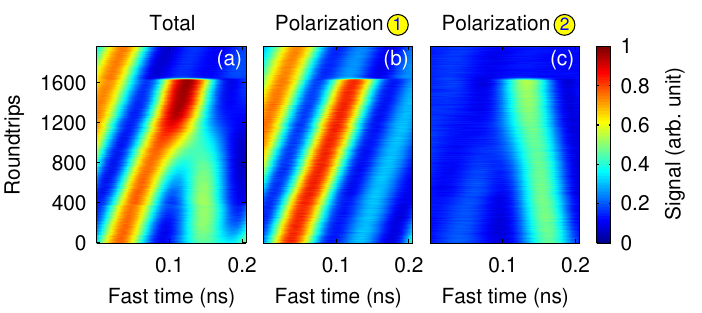}
		\caption{Measurements showing the collision and annihilation of two different CSs. (a) Total intensity; (b) intensity along polarization mode 1; (c) intensity along polarization mode 2. The CS to the far left (in mode 1) does not play a role in the collision.}
		\label{fig:4}
\end{figure}

\begin{figure}[t]	
		\includegraphics[width=\linewidth]{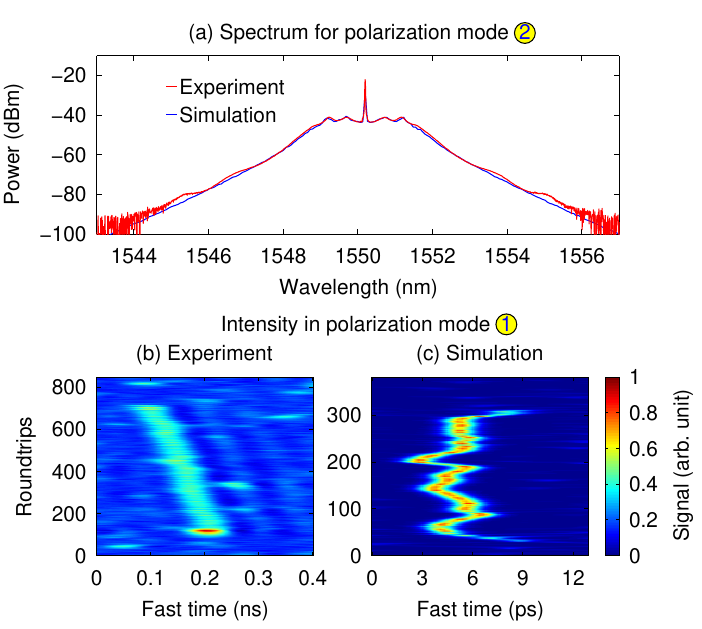}
		\caption{Observation of spontaneous creation and destruction of a CS by an unstable MI pattern. (a) Measured and simulated spectrum for mode 2. The smoothness is indicative of a fully unstable MI pattern. (b, c) Experimentally measured (b) and simulated (c) dynamics showing the spontaneous appearance and disappearance of a CS in mode 1. The erratic CS motion observed in simulations occurs over time-scales that cannot be resolved in experiment [note the different axes in (b) and (c)].}
		\label{fig:5}
\end{figure}

During transients, where the CSs are not yet pinned to amplitude inhomogeneities, the group-velocity mismatch can lead to collisions and other forms of interactions between coexisting CSs. Our experiments show a rich diversity of possible interaction scenarios, ranging from the formation of bound states (as predicted in~\cite{suzuki_theoretical_2018}) to soliton annihilation. Figure~\ref{fig:4} shows an example of a temporally resolved experimental measurement of CS evolution dynamics. Here, Fig.~\ref{fig:4}(a) shows the total intracavity intensity of two nonidentical solitons colliding and annihilating each other. Polarization resolved measurements shown in Figs.~\ref{fig:4}(b) and (c) clearly reveal that the solitons are associated with different polarization modes. Similar dynamics can be reproduced in our simulations when adding to Eqs.~\eqref{eq:1} an additional drift term that represents the group-velocity mismatch between the polarization modes.

It is interesting to note that we have been unable to observe, both experimentally and numerically, coexistence between CSs and fully unstable MI patterns with a characteristically smooth spectrum. [The spectrum in Fig.~\ref{fig:3}(d) shows significant residual structure.] We speculate this is because large power spikes in the MI pattern destroy any CSs in polarization mode \circled{1} through XPM. Conversely, such a spike perturbing the cw steady-state in polarization mode \circled{1} could also be envisaged to spontaneously excite CSs in polarization mode \circled{1}. We have not, however, observed such spontaneous excitation for the parameters considered above, possibly due to the weakness of the MI fluctuations in polarization mode \circled{2}. To strengthen the fluctuations and hence test the underlying hypothesis, we shifted the two resonances closer to each other ($\delta\Delta = 12.5$), balanced the driving ($\chi = \frac{\pi}{4}$) and fixed the detuning at a value of $\Delta_{1} = 14.8$. For these parameters, mode \circled{2} displays a smooth spectrum indicative of a fully unstable MI field [Fig.~\ref{fig:5}(a)], while the spectrum of mode \circled{1} (not shown) is void of any CS-like features. Yet remarkably, time resolved measurements show that CSs can intermittently emerge from the cw background. Figure~\ref{fig:5}(b) shows an example of such a scenario: a CS can be seen to spontaneously appear and persists for hundreds of round trips before again spontaneously disappearing. Numerical simulations qualitatively reproduce the behaviour, with an example scenario shown in Fig.~\ref{fig:5}(c).

In conclusion, we have reported on experimental observations of coexistence and interactions between nonlinear states with different polarizations in a monochromatically driven passive Kerr resonator. In addition to experimentally confirming earlier theoretical predictions~\cite{averlant_coexistence_2017,suzuki_theoretical_2018}, our work shows that the simultaneous excitation of several cavity modes with a  single carrier frequency can engender very rich dissipative dynamics. The ability to control the frequency separation between two modes (as demonstrated in our work) paves the way for further studies of such dynamics, and will allow for the systematic exploration of questions such as: what are the exact parameter conditions required for coexistence of nonlinear states; how does linear mode coupling affect the cavity dynamics; can the coexistence of nonlinear states be linked to the well-known symmetry breaking behaviour of driven Kerr resonators~\cite{haelterman_polarization_1994,copie_interplay_2019}? In addition to expanding our fundamental knowledge of dissipative solitons and multimode Kerr resonators, answering questions such as these could impact on the design of single-source multi-output microresonator frequency combs.

\begin{acknowledgments}
The authors wish to acknowledge financial support from the Marsden Fund and the Rutherford Discovery Fellowships of the Royal Society of New Zealand.
\end{acknowledgments}

\end{document}